\documentclass[aps,prb,twocolumn,groupedaddress]{revtex4}

\usepackage{graphics} 
\usepackage{graphicx}
\usepackage{dcolumn}
\usepackage{bm}
\everymath{\displaystyle}
\usepackage{amsmath}

\setcitestyle{super}

\begin{document}

\title{Anomalous Decay of Quantum Resistance Oscillations \\of  Two Dimensional Helical Electrons in Magnetic Field.} 

\author{S. Abedi}
\author{S. A. Vitkalov}
\email[Corresponding author: ]{svitkalov@ccny.cuny.edu}
\affiliation{Physics Department, City College of the City University of New York, New York 10031, USA}
\author{N. N. Mikhailov}
\author{Z. D. Kvon}
\affiliation{A.V.Rzhanov Institute of Semiconductor Physics, Novosibirsk 630090, Russia}
\affiliation{Novosibirsk State University, 630090 Novosibirsk, Russia}

\date{\today}

\begin{abstract} 
 Shubnikov de Haas (SdH) resistance oscillations of highly mobile two  dimensional helical electrons propagating on a conducting surface of strained HgTe 3D topological insulator are studied in magnetic fields $B$ tilted by angle $\theta$ from the normal  to the conducting layer.  Strong decrease of    oscillation amplitude $A$ is observed with the tilt: $A \sim \exp(-\xi/cos(\theta))$, where $\xi$ is a constant. Evolution  of the oscillations with temperature $T$ shows that the parameter $\xi$ contains two terms: $\xi=\xi_1+\xi_2 T$. The temperature independent term, $\xi_1$, describes reduction of  electron mean free path $l_q$ in magnetic field $B$ pointing toward  suppression of the topological protection of the electron states against impurity scattering.  The temperature dependent term, $\xi_2 T$, indicates  increase of  the reciprocal velocity  of 2D helical electrons : $\delta (v_F^{-1})$$\sim$$B$ suggesting  modification of the electron spectrum in magnetic fields.

\end{abstract}
 

\maketitle

Two- and three-dimensional topological insulators (3D TIs) represent a new class of materials with an insulating bulk and topologically protected conducting boundary states.\cite{kane2005a,kane2005b,bernevig2006a,bernevig2006b,Fu2007a,Fu2007b,hsien2008,hasan2010,zhang2011,ando2013} In 3D TIs, due to a strong spin-orbit interaction,  a propagating surface  electron state with  wave vector ${\bf k}$ is non-degenerate and keeps the electron spin locked perpendicular to the  wave vector ${\bf k}$ in the 2D plane (2D helical electrons).\cite{Fu2007a,zhang2011,ando2013} Due to the spin-momentum  locking, the electron scattering on impurities is suppressed since the scattered electron should change both the linear and the angular (spin)  momenta. It leads to  a topological protection of the helical electrons against the scattering. In particular, the 180$^o$ backscattering is  expected to be absent\cite{hasan2010,zhang2011,ando2013}.   The topological protection is predicted to enhance the mobility of helical electrons and is the reason why TIs are considered for various applications.\cite{moore2010}  

A predicted 3D topological insulator, based on strained HgTe films,\cite{Fu2007a} has been recently realized\cite{brune2011,brune2014}   and a very high mobility (approaching 100 m$^2$/Vs) of 2D helical electrons in this system is achieved.\cite{kozlov2014,kozlov2016}   The high mobility facilitates  measurements of  transport properties, in particular,  Landau quantization of helical electrons  down to low magnetic fields \cite{brune2011,kozlov2014,brune2014,kozlov2016} and has provided a direct transport verification of the non-degeneracy of the helical surface states in strained HgTe films.\cite{maier2017} One of the reasons of the high mobility is  well developed technology of the fabrication of  HgTe films with a low  density of impurities.  Another reason might  be the topological protection of the helical electron states against the impurity scattering.\cite{hasan2010,zhang2011,ando2013} This protection  is scarcely seen in  transport measurements due to a low electron mobility (below 1 m$^2$/Vs)  in the majority of 3D TI materials.\cite{taskin2012}  A magnetic field breaks the time reversal symmetry responsible for the lack of the backscattering\cite{hasan2010,zhang2011,ando2013} and increases the spin overlap between incident and scattered electron states. Thus, the magnetic field  should  increase the impurity scattering  of the 2D helical electrons.  To the best of our knowledge such magnetic field induced enhancement of the scattering of 2D helical electrons has not been  reported yet. 

Below we present  transport investigations  of quantum resistance oscillations of highly mobile 2D helical electrons in HgTe strained films placed in tilted magnetic fields. Due to  the spin-momentum locking  a propagating quantum  state of  a 2D helical electron is non-degenerate  and, thus, cannot split in a magnetic field. In contrast  the spin degenerate propagating state of an ordinary 2D electron splits on spin-up and spin-down levels by the magnetic field that leads to large variations of the amplitude of SdH oscillations in tilted magnetic fields\cite{fang1968,ando1982}.    Thus,  the angular variations of SdH resistance oscillations of 2D helical electrons   are not expected since the electron spin  non-degenerate quantum states do not split.

Surprisingly, the experiments show that, despite  the spin non-degeneracy of the electron spectrum, the magnetic field reduces strongly the amplitude of the quantum oscillations.  A comprehensive investigation of this effect shows that a quantum mean free path $l_q$ of the 2D helical electrons, which at low temperatures is controlled  by the impurity scattering,  decreases significantly with the magnetic field. We relate this decrease to the  magnetic field suppression of the topological protection of the 2D helical electrons against the impurity scattering. Furthermore, an analysis of the  evolution of the  oscillation amplitude with the  temperature indicates a linear increase of the reciprocal Fermi velocity $v_F^{-1}$ of 2D helical electrons with the  magnetic field:  $ \delta (v_F^{-1})\sim B$.  This effect  suggests a  modification of   the dynamics of 2D helical electrons in magnetic fields. 
         
\begin{figure}[t]
\vskip -0 cm
\includegraphics[width=\columnwidth]{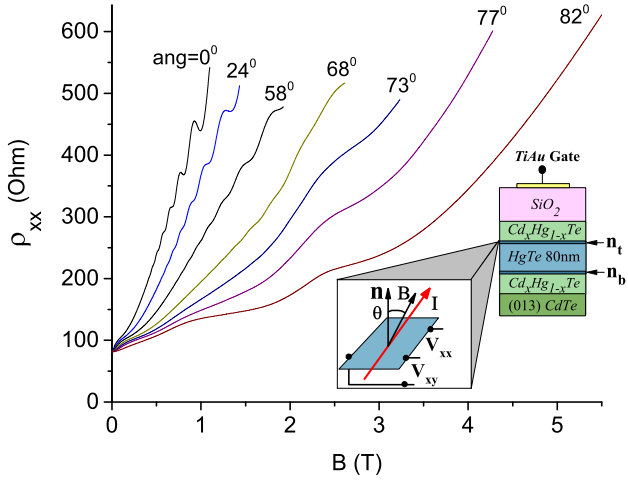}
\caption{(color online) Dependence of  resistivity $\rho_{xx}$ of 2D helical electrons on magnetic field, $B$, applied at different angles $\theta$ with respect to  HgTe layers as labeled. Visible at $\theta$=0$^0$ oscillating content disappears at $\theta>$ 73$^0$. The insert shows the studied structures and  geometry of the experiments.  Sample TI5. $V_g$=2.5V. T=4.2K.  
}
\label{fig1}
\end{figure}

\begin{figure}[t]
\vskip -0 cm
\includegraphics[width=\columnwidth]{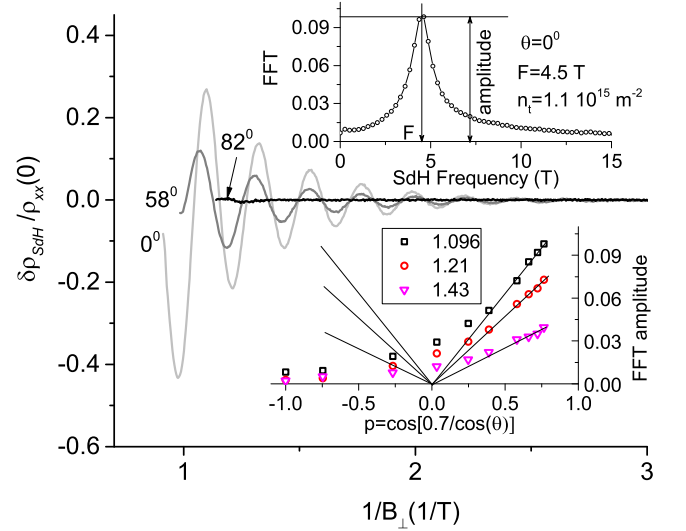}
\caption{(color online) Dependence of  normalized resistance oscillations  $\delta \rho_{SdH}/\rho_{xx}(0)$ of 2D helical electrons on reciprocal perpendicular magnetic field, $B_\perp^{-1}$,  at different angles $\theta$ as labeled. The amplitude of the SdH oscillations reduces with the angle and is zero at $\theta$=82$^0$.   Upper insert  shows FFT spectrum of the oscillations started at $(B_\perp^{-1})^L$=1.09(1/T) at $\theta$=0$^0$. Lower insert shows an angular dependence of FFT amplitude indicating  significant disagreement with the one expected for ordinary degenerate 2D electrons and shown by solid straight lines.  Sample TI5. $V_g$=2.5V. T=4.2K.  
}
\label{fig2}
\end{figure}

Studied, 80 nm wide, strained HgTe films are grown by molecular beam epitaxy on (0,1,3) CdTe substrate.  Since HgTe films grown directly on CdTe suffer from dislocations due to the lattice mismatch, our 80 nm thick HgTe films were separated from the CdTe substrate by a 20 nm thin Cd$_{0.7}$Hg$_{0.3}$Te buffer layer. This buffer layer significantly increases  the electron mobility up to 40 m$^2$/(Vs).\cite{kozlov2014} In Fig.\ref{fig1}   the insert  shows the studied structures. The 2D helical electrons are formed at the top and the bottom surfaces of the HgTe film. The structures are equipped with a TiAu gate providing the possibility to tune the Fermi  energy $E_F$ inside  the insulating gap $\Delta_g \approx$15 mV\cite{kozlov2014,gap} and to change  the density $n=n_t+n_b$ of 2D helical electrons, where  $n_t$ ($n_b$) is the density of 2D electrons located at the top (bottom) of HgTe film.  Magnetotransport experiments indicate that  at a positive gate voltage $V_g$,   $n_t>n_b$ since   the top HgTe surface is closer to the gate.\cite{kozlov2014}  

Samples are etched in the shape of a Hall bar with width  $W=50\mu$m.  Two samples are studied in magnetic fields up to 8 Tesla applied  at different angle $\theta$  relative  to the normal ${\bf n}$ to 2D layers and  perpendicular to the applied current.  The angle $\theta$ is evaluated using Hall resistance $R_{xy}$, which is proportional to the perpendicular component, $B_\perp=B cos(\theta)$, of the total magnetic field $B$.     Experiments indicate that 2D helical electrons located at the top of HgTe film provide the dominant contribution to SdH oscillations at small magnetic fields.\cite{kozlov2014,kozlov2016} The density $n_t$ is estimated  from  the frequency of SdH oscillations  taken at $\theta$=0$^0$ (see upper insert to Fig.\ref{fig2}).  An averaged mobility obtained from Hall resistance and the resistivity at zero magnetic field for sample TI1 (TI5) is $\mu$=43 m$^2$/Vs (37m$^2$/Vs). Sample resistance was measured using the four-point probe method. We applied a 133 Hz $ac$ excitation $I_{ac}$=0.5$\mu$A  through the current contacts and measured the longitudinal (in the direction of the electric current, $x$-direction) and Hall (along $y$-direction) voltages.  The measurements were done in the linear regime in which the voltages are proportional to the applied current. 

Figure \ref{fig1} shows the dissipative magnetoresistivity  $\rho_{xx}(B)$ taken at different angles $\theta$ as labeled.  Quantum resistance oscillations are visible at $\theta$=0$^0$, 24$^0$ and 58$^0$ and are  significantly suppressed at $\theta>$68$^0$. To facilitate an analysis of the oscillating content, the monotonic background, obtained by an adjacent point averaging over the period of the oscillations in  reciprocal magnetic fields, is removed from the magnetoresistivity $\rho_{xx}(B)$.  Figure \ref{fig2} presents the remaining oscillating content of the magnetoresistivity, $\delta \rho_{SdH}$,  normalized by $\rho_{xx}(B=0)$ as a function of the reciprocal perpendicular magnetic field $B_\perp^{-1}$.\cite{ratio} As expected, the SdH oscillations are periodic in $B_\perp^{-1}$.\cite{shoenberg1984,ando1982} In agreement with Fig.\ref{fig1}, SdH oscillations decrease with the angle $\theta$ and are absent at $\theta$=82$^0$. The upper insert shows the Fourier spectrum  obtained by  Fast Fourier Transformation (FFT) of the oscillations between $1/B_\perp^L$=1.09 (1/T) and  $1/B_\perp^R$=5 (1/T) at $\theta$=0$^0$.\cite{fft} The SdH frequency $F$=4.5(T) yields the 2D electron density $n_t=(e/h) F$=1.1 10$^{15}$ m$^{-2}$.\cite{shoenberg1984,ando1982} The density $n_t$ stays the same at  different  angles $\theta$.  A comparison with the Hall coefficient indicates a presence of a second group of 2D electrons with a density $n_b$=0.7 10$^{15}$ m$^{-2}$, which should oscillate at frequency 2.9(T). These oscillations are absent in the spectrum at small $B_\perp$, which is  consistent with previous experiments.\cite{kozlov2014}

\begin{figure}[t]
\vskip -0 cm
\includegraphics[width=\columnwidth]{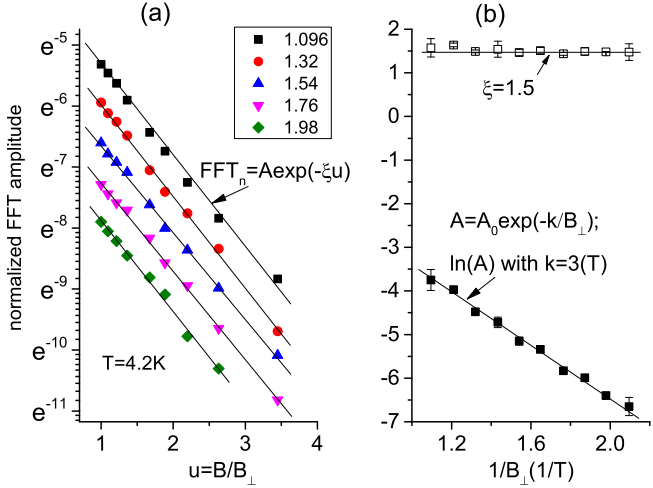}
\caption{(color online) (a) Dependence of  normalized FFT amplitude   on $B/B_\perp$.  FFT amplitude is obtained for SdH oscillations in  interval [$B^{-1}_\perp$,5] T$^{-1}$. Symbols correspond to different $B^{-1}_\perp$ as labeled. The dependence is fitted in accordance with Eq.(\ref{fft}).  (b) Dependence of parameters $\xi$ and A extracted from the fit on $B^{-1}_\perp$. The parameter $\xi$=1.5$\pm$0.15 indicates uniform ($B_\perp$-independent) relative decrease of SdH amplitude with angle $\theta$.  Sample TI5. $V_g$=2.5V. T=4.2K.  
}
\label{fig3}
\end{figure}

The lower insert shows a comparison of  the angular dependence of FFT amplitude with the one expected for spin degenerate 2D electron states and shown by the straight solid lines. The Zeeman effect splits the spin degenerate electron quantum levels leading to a variation of the SdH amplitude  with the angle\cite{fang1968,mayer2016}: $A\sim p=\cos(\pi \Delta_Z/\Delta_c)=\cos(\gamma/\cos(\theta))$ since the cyclotron energy $\Delta_c \sim B_\perp$ and the Zeeman energy $\Delta_Z \sim B$.  The fitting parameter $\gamma=\pi g m_c/m_0$=0.7, where $g$ is electron g-factor and $m_0$ is free electron mass,  is chosen to provide the best fit  with the experiment. The comparison  yields g-factor $g\approx$12 at cyclotron mass $m_c \approx$0.02$m_0$ determined from temperature experiments presented below. The discrepancy between the experiment and the  behavior of the spin degenerate 2D electrons is seen at $p<$0.4, when the experimental data  deviates from the straight lines. At $p<$0, the SdH amplitude continues to decrease in  contrast to the spin degenerate case.    

To analyze the observed decrease of the amplitude of SdH oscillations in a spin non-degenerate electron system, such as 2D helical electrons in the strained HgTe films,\cite{Fu2007a,brune2011,kozlov2014,brune2014,kozlov2016,maier2017} one should assume that some physical parameters, controlling the SdH amplitude  in  Lifshits-Kosevich formula\cite{shoenberg1984,ando1982},  change with the magnetic field. The following relations  of the quantum mean free path $l_q$ and Fermi velocity $v_F$ with  the magnetic field B: $l_q^{-1}=l_0^{-1}(1+\alpha B) $ and   $v_F^{-1}=v_0^{-1}(1+\beta B)$, where $l_0,v_0, \alpha, \beta$ are constants,  reproduce the observed results. A substitution of these relations   into Lifshits-Kosevich formula  at small magnetic fields\cite{ando1982,mayer2016}  yields  a normalized FFT amplitude\cite{norm}:
 \begin{figure}[t]
\vskip -0 cm
\includegraphics[width=\columnwidth]{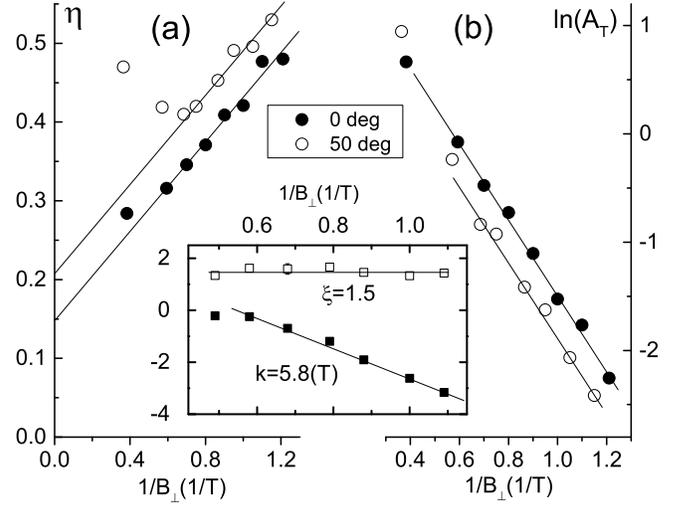}
\caption{ (a) Dependence of  parameter $\eta=\beta a u+a/B_\perp$  on $B^{-1}_\perp$. The parameter is extracted  from   FFT$_n$(T,u=const)  set obtained at different  temperatures T in the interval 5.5 to 12.5 K at a fixed angle similar to the parameter $\xi$ extracted from the FFT$_n$(u,T=const) set obtained at fixed temperature and shown in Fig.\ref{fig3}; (b) Dependence of amplitude $A_T$ on $B^{-1}_\perp$. Different symbols correspond to different angles $\theta$ as labeled. Insert shows result of the rotation experiment at  fixed temperature T=5.5K for the same sample. Sample TI1. $V_g$=1.7V. $n_t$=1.2 10$^{15}$ m$^{-2}$. 
}
\label{fig4}
\end{figure}      
\begin{equation}
FFT_n(u,T,B_\perp)=A_0\exp(-\xi u) \exp(-\frac{k}{B_\perp})    
\label{fft}
\end{equation}   
Here $\xi=\alpha d+\beta a T$, $k=d+aT$, $d=\pi \hbar k_F/(el_0)$, $a=2\pi^2k_B k_F/(ev_0)$, $u$=$B/B_\perp$=$1/\cos(\theta)$, $k_F$ is the electron wave number at $E_F$, $k_B$ is Boltzmann constant, $e$ is electron charge  and $A_0$ is a constant.  Parameters $d$ and $a$ are coming from  the Dingle and temperature damping factors of the SdH amplitude.\cite{ando1982,mayer2016} A relation $\omega_c=ev_F B_\perp/(\hbar k_F)$ is used for the cyclotron frequency. 

Figure \ref{fig3}(a) shows a dependence of the normalized amplitude FFT$_n$ on $u=B/B_\perp$ at different $B_\perp$ as labeled. In a broad range of $B_\perp$, the SdH amplitude decreases exponentially with $u$. Figure \ref{fig3}(b) shows that the parameter $\xi$ is nearly independent on $B_\perp$, while the SdH magnitude $A$ drops exponentially with  $1/B_\perp$.  Similar results are obtained at different densities $n_t$ on both samples. 

Measurements at different temperatures  indicate  presence of a temperature dependent contribution to $\xi$.\cite{tdep}  Figure \ref{fig4}(a) shows  a dependence of the parameter $\eta=\beta a u+a/B_\perp$  on $1/B_\perp$. This parameter controls the exponential temperature variations of FFT$_n$ in Eq.(\ref{fft}). The presented parameter $\eta$  is  obtained  from the T dependence of FFT$_n$(T,u=const)   similar to the parameter $\xi$ found from  FFT$_n$(u,T=const) shown in Fig.\ref{fig3}. Fig.\ref{fig4}(a) demonstrates  that the parameter $\eta$ decreases linearly with decreasing $1/B_\perp$ as expected\cite{ando1982} yielding $a$=0.28$\pm$0.03(T/K) and the velocity $v_0$= 7.5($\pm 0.8$)10$^5$m/s.  However, in  contrast to  the ordinary 2D electrons, the parameter  $\eta$  does not extrapolate to zero at $1/B_\perp$$\rightarrow$0, indicating  a non-zero term $\eta_0=\beta a u$=0.15$\pm 0.03$,  yielding $\beta$=0.5$\pm$0.15 at u=1 ($\theta$=0$^0$).  Taken at different   angle $\theta$=50$^0$ measurements show a consistent increase of the $\eta_0$ with the angle: $\eta_0(u=1.54)$=0.21$\pm$0.03.\cite{botlay}

Figure \ref{fig4}(b) presents a behavior of  FFT amplitude $A_T=\exp[-(\alpha d u+ d/B_\perp)]$  obtained from the same T-dependence of FFT$_n$(T,u=const) (similar to A in Fig.\ref{fig3}(b)). The slope of the linear dependence $\ln(A_T)$ vs. $1/B_\perp$ yields $d$=3.5$\pm0.3$. Thus at density $n_t$=1.2 10$^{15}$ m$^{-2}$ the quantum mean free path is $l_0$=73 nm in the studied sample.  At $\theta$=50$^0$(u=1.54), the dependence shifts down yielding $\xi_1$=$\alpha d$=0.76$\pm$0.15 and $\alpha$=0.22$\pm$0.05. Using the  obtained parameters $a,d,\alpha, \beta$   we evaluate the parameters $\xi_{ev}$=1.58$\pm$0.35 and $k_{ev}$=5$\pm$0.8. The estimated parameters are close to the ones obtained in the rotation experiments: $\xi$=1.5$\pm$0.1 and $k$=5.8$\pm$0.3, and  shown in the insert to Fig.\ref{fig4}.  Thus, the cross examination indicates a consistency of the obtained results. A similar outcome is found at different electron density $n_t$=1.6 10$^{15}$ m$^{-2}$ (not shown).  
\begin{figure}[t]
\vskip -0 cm
\includegraphics[width=\columnwidth]{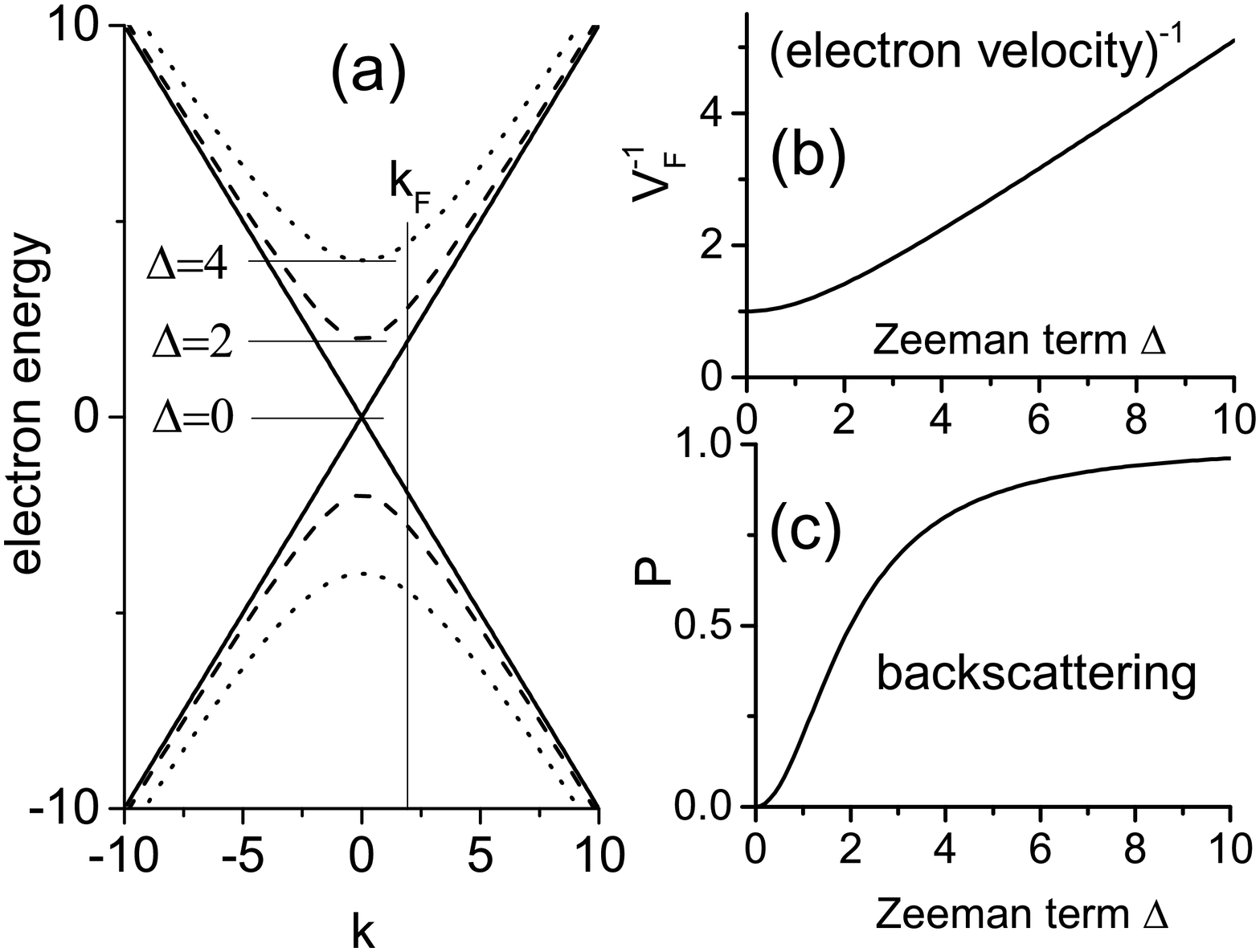}
\caption{ (a) counted from $C$ energy dispersion $\epsilon({\bf k })$ of 2D helical electrons at different values of  Zeeman energy $\Delta$ as labeled;  (b) variations of reciprocal Fermi velocity with Zeeman energy; (c) variations of normalized probability $P$ of electron backscattering  with Zeeman energy. $k_F$=2. All parameters are in  relative units.  
}
\label{fig5}
\end{figure}

Fig.\ref{fig3} and Fig.\ref{fig4} show good agreement between the experiements and Eq.(\ref{fft})   revealing  unexpected and strong  suppression of SdH oscillations of 2D helical electrons with the magnetic field $B$. At a fixed $B_\perp$ the amplitude of SdH oscillations decreases exponentially with $B$: $A_{SdH} \sim \exp[-(\xi/B_\perp)B]$  indicating  possible relevance of a spin effect, which is proportional to $B$.   

In response to the Lorentz force, ${\bf F}_L=e{\bf v}\times {\bf B}$, electrons in a single band move  in accordance with the quasi-classical theory, considering effects of the Lorentz force on the band structure to be negligibly small.\cite{ziman} In the systems with no spin-orbit interaction the ${\bf k}$-space and spin ${\bf s}$-space are disentangled. A change of the electron energy via  Zeeman effect repopulates the spin-up and spin-down subbands in the ${\bf k}$-space keeping the energy dispersion of electrons  intact:  $\epsilon_\uparrow({\bf k })$=$\epsilon_\downarrow({\bf k })$. Thus at a fixed $k_F$ (electron density)   both the Lorentz force and Zeeman effect should not change the Fermi velocity  $v_F$.   In systems  with a spin-orbit coupling a variation in the ${\bf s}$-space via the Zeeman term,  may change the electron dispersion in the ${\bf k}$-space and lead to a variation of the electron velocity $v_F$. To illustrate this effect  we consider a simple model of 2D helical electrons affected by the Zeeman term $\Delta \sim {\bf B}=(0,0,B_z)$ .   The following Hamiltonian describes  2D helical states of a 3D topological insulator (see Eq.(34) in Ref.[9])\cite{liu2010}:   
 \begin{equation}
H=C+A(\sigma^x k_y-\sigma^y k_x)+\Delta \sigma^z
\label{ham}
\end{equation}  
where C and A  material constants, $\sigma^{x,y,z}$ are Pauli matrices and ${\bf k}=(k_x, k_y)$ in the 2D electron wave vector. The Zeeman term $\Delta \sigma^z$  changes the electron spectrum leading to a spectral gap:
 \begin{equation}
\epsilon(k)=C \pm ( \Delta^2+A^2k^2)^{1/2}
\label{energy}
\end{equation}    
 Fig.\ref{fig5}(a) presents the electron spectrum at different strengths of the Zeeman term as labeled. The vertical thin line indicates the electron wave number $k_F$ at Fermi energy.  Fig.\ref{fig5}(b) shows the increase of the reciprocal Fermi velocity $v_F^{-1}=(\partial \epsilon /\partial k)^{-1}$($k$=$k_F$) with  $\Delta$, following from Eq.(\ref{energy}). The increase  is proportional to $B$ at a large $\Delta$.
 The model also shows an increase of the electron scattering in magnetic fields. By polarizing electron spins in the $z$-direction the magnetic field increases the spin overlap between incident ${\bf k}_F(\theta_{in})$ and scattered ${\bf k}_F(\theta_{fin})$ electron states. Fig.\ref{fig5}(c) presents the dependence of  a normalized rate of the electron backscattering ($\theta_{fin}-\theta_{in}=\pi)$    on  the Zeeman term $\Delta \sim B$.\cite{scalar} At $\Delta=0$ the rate is zero indicating  the topological protection of the backscattering. With  increasing $\Delta$ the  rate increases imitating  a linear dependence on $\Delta$ in the interval from 0.5 to 2.  At high $\Delta$ the rate approaches  1 indicating that at high magnetic fields there are no spin restrictions on the impurity scattering  since all electron spins are polarized along ${\bf B}$.   

Despite the similarity with the experiment  the  model  is not directly applicable to the presented data. The studied 2D helical electrons are a result of a linear superposition of electron states  from several subbands and additional terms may affect the spectrum.\cite{raichev2012}    Further investigations are required to explain the presented findings quantitatively and reveal the dominant mechanism (s) leading to the anomalous angular decay of quantum resistance oscillations of 2D helical electrons.    
  
In summary, the angular dependence  of quantum resistance oscillations of 2D helical electrons  in   3D topological insulators based on strained HgTe films demonstrates exponentially strong reduction of the oscillation amplitude $A$ with the magnetic field $B$:  $A \sim \exp[-(\xi/B_\perp)B]$. The  temperature dependence  of the amplitude $A$  exhibits two terms contributing  to the parameter $\xi$:  $\xi=\xi_1+\xi_2 T$.  The temperature independent term, $\xi_1$  indicates considerable reduction of the quantum mean free path $l_q$ in the magnetic field $B$. The reduction is  consistent with the form: $[\delta(l_q^{-1})]/l_0^{-1}$=$\alpha B$, where $\alpha$=0.22$\pm$0.03(T$^{-1}$) at electron density $n_t$=1.2 10$^{15}$ m$^{-2}$. The $l_q$ decrease is related to the  suppression of the topological protection of the helical electron states against the  impurity scattering in magnetic fields.  The temperature dependent term, $\xi_2 T$, indicates significant  increase  of  the reciprocal velocity $v_F^{-1}$ of 2D helical electrons in the magnetic field, which is  consistent with the form:  $[\delta v_F^{-1}]/v_0^{-1}=\beta B$, where $\beta$=0.5$\pm$0.15(T$^{-1}$) at $n_t$=1.2 10$^{15}$ m$^{-2}$. This increase suggests that the  magnetic field considerably modifies  the dynamics of  2D helical electrons.   

S. V. thanks Prof. I. Aleiner for useful discussion.This work was supported by the National Science Foundation (Division of Material Research -1702594).  Novosibirsk team is supported by Russian Science Foundation (Grant No. 16-12-10041).


\begin{thebibliography} {}

\bibitem{kane2005a}C. L. Kane and E. J. Mele, Phys.Rev.Lett. {\bf 95}, 146802 (2005).
\bibitem{kane2005b} C. L. Kane and E. J. Mele, Phys. Rev. Lett. 95, 226801 (2005).
\bibitem{bernevig2006a} B. A. Bernevig and S. -C. Zhang, Phys. Rev. Lett. 96, 106802 (2006).
\bibitem{bernevig2006b} B. A. Bernevig, T. L. Hughes, and S.-C. Zhang, Science 314, 1757 (2006).
\bibitem{Fu2007a} Liang Fu and C. L. Kane, Phys. Rev. B {\bf 76}, 045302 (2007).
\bibitem{Fu2007b} L. Fu, C. L. Kane, and E. J. Mele, Phys. Rev. Lett. 98, 106803 (2007).
\bibitem{hsien2008} D. Hsieh, D. Qian, L. Wray, Y. Xia, Y. S. Hor, R. J. Cava, and M. Z. Hasan, Nature (London) 452, 970 (2008).
\bibitem{hasan2010} M. Z. Hasan and C. L. Kane, Rev. Mod. Phys. 82, 3045 (2010).
\bibitem{zhang2011} X. -L. Qi and S.-C. Zhang, Rev. Mod. Phys. 83, 1057 (2011).
\bibitem{ando2013} Y. Ando, J. Phys. Soc. Jpn. 82, 102001 (2013).
\bibitem{moore2010} J. E. Moore, Nature (London) 464, 194 (2010).


\bibitem{brune2011} C. Brune, C.X. Liu, E.G. Novik, E.M. Hankiewicz, H. Buhmann, Y.L. Chen, X.L. Qi, Z.X. Shen, S.C. Zhang, and L.W. Molenkamp, Phys. Rev. Lett. 106, 126803 (2011).
\bibitem{kozlov2014} D. A. Kozlov, Z. D. Kvon, E. B. Olshanetsky, N. N.
Mikhailov, S. A. Dvoretsky, and D. Weiss, Phys. Rev. Lett. 112, 196801 (2014).
\bibitem{brune2014}C. Brune, C. Thienel, M. Stuiber, J. Bottcher, H. Buhmann,
E. G. Novik, C.-X. Liu, E. M. Hankiewicz, and L.W. Molenkamp, Phys. Rev. X 4, 041045 (2014).
\bibitem{kozlov2016} D. A. Kozlov, D. Bauer, J. Ziegler, R. Fischer, M. L. Savchenko, Z. D. Kvon,
N. N. Mikhailov, S. A. Dvoretsky, and D. Weiss, Phys. Rev. Lett. 116, 166802 (2016).
\bibitem{maier2017} Hubert Maier, Johannes Ziegler, Ralf Fischer, Dmitriy Kozlov, Ze Don Kvon, Nikolay Mikhailov,Sergey A. Dvoretsky and  Dieter Weiss, Nature Comm. {\bf 8}, 2023, (2017)


\bibitem{taskin2012} A. A. Taskin, Satoshi Sasaki, Kouji Segawa, and Yoichi Ando, Phys. Rev. Lett. {\bf 109}, 066803 (2012).

\bibitem{fang1968}F. F. Fang,  and P. J. Stiles, Phys. Rev. {\bf 174}, 823 (1968).
\bibitem{ando1982} T. Ando, A. B. Fowler, and F. Stern, Rev. of Mod. Phys. B {\bf 54}, 437 (1982).

\bibitem{gap} Reported in this paper measurements are done,  when Fermi energy is inside the gap $\Delta_g$.

\bibitem{ratio} The presented data are obtained when $\rho_{xy} \gg \rho_{xx}$. In this regime $\rho_{xx}/\rho_{xx}(B=0)\approx \sigma_{xx}/\sigma_D$ and the amplitude of quantum oscillations in the resistivity is  proportional to the one in the conductivity $\sigma_{xx}$  with an accuracy better than 3 percent in the experiemts. Here $\sigma_D$ is Drude conductivity.

\bibitem{shoenberg1984} D. Shoenberg {\it Magnetic oscillations in metals}, (Cambridge University Press, New York, 1984). 

\bibitem{fft} FFT analysis enhances the  separation of the SdH oscillations of 2D electrons located at the top surface from the SdH response of the bottom conducting surface oscillating at a different frequency.   

\bibitem{mayer2016} William Mayer, Areg Ghazaryan, Pouyan Ghaemi, Sergey Vitkalov A. A. Bykov, Phys. Rev. B {\bf 94}, 195312 (2016).

\bibitem{norm} $FFT_n(u,T,B_\perp^L)=FFT(u,T,B_\perp^L)/C(u,T,B_\perp^L)$, where the  factor   $C(u,T,B_\perp^L)=4aT[(d+aT)(3\beta u+1/B^L_\perp)+1]/(d+aT)^2$. In practice the polynomial factor $C(u,T,B_\perp^L)$ uses parameters $a,d,\beta$ obtained self-consistently with  the exponential fits  shown in Fig.(\ref{fig4})  via several  iterations.

\bibitem{tdep} Estimations indicate that contributions of the inelastic processes (such as electron-electron , electron-phonon scattering) to the observed temperature dependence of parameter $\xi$ are small. These processes are ignored in the paper.    

\bibitem{botlay}  In Fig.\ref{fig4}(a)  deviations from the  straight lines  at $B^{-1}_\perp <$0.5 T$^{-1}$ are related to an interference between SdH oscillations of 2D electrons located at the top and  bottom surfaces of HgTe film. Contributions from the bottom 2D layer oscillate at frequency $F_b \approx$ 3 T and  are visible in the  FFT spectrum at small $B^{-1}_\perp$ in agreement with the previous study.\cite{kozlov2014}   

\bibitem{liu2010} Liu, C.-X., X.-L. Qi, H. Zhang, X. Dai, Z. Fang, and S.-C. Zhang, Phys. Rev. B {\bf 82}, 045122 (2010). 
\bibitem{scalar} The presented probability  is a square of the magnitude of the  scalar product of two eigenvectors of Hamiltonian (\ref{ham}) corresponding to  incident ${\bf k}_F(\theta_{in})$ and scattered ${\bf k}_F(\theta_{fin})$ electron states.

\bibitem{ziman}J. M. Ziman {\it Principles of the theory of solids}, (Cambridge at the University Press, 1972).
\bibitem{raichev2012} O. E. Raichev, Phys. Rev. B {\bf 85}, 045310 (2012).



\end{thebibliography}
\end{document}